\documentclass[aps,twocolumn,showpacs,preprintnumbers,pra]{revtex4}

\usepackage{graphicx}
\usepackage{dcolumn}
\usepackage{bm}

\begin{document}

\preprint{paper1.tex}

\title{Theory of inelastic lifetimes of surface-state electrons and holes at
metal surfaces}
\author{J. M. Pitarke$^{1,2}$ and M. G. Vergniory$^{2,3}$}
\affiliation{
$^1$CIC nanoGUNE Consolider, Mikeletegi Pasealekua 56, E-20009 Donostia, Basque
Country\\
$^2$Materia Kondentsatuaren Fisika Saila, UPV/EHU, and Centro F\'\i sica
Materiales CSIC-UPV/EHU,\\ 644 Posta kutxatila, E-48080 Bilbo, Basque Country\\
$^3$Donostia International Physics Center (DIPC),\\
Manuel de Lardizabal Pasealekua, E-20018 Donostia, Basque Country}

\date{\today}

\begin{abstract}
After the early suggestion by John Pendry to probe unoccupied bands at surfaces
through the time reversal of the photoemission process, the
inverse-photoemission technique yielded the first conclusive experimental
evidence for the existence of image-potential bound states at metal surfaces
and has led over the last two decades to an active area of research in
condensed-matter and surface physics. Here we describe the current status of
the many-body theory of inelastic lifetimes of these image-potential states
and also the Shockley surface states that exist near the Fermi level in the
projected bulk band gap of simple and noble metals. New calculations of the
self-energy and lifetime of surface states on Au surfaces are presented as
well, by using the $GW\Gamma$ approximation of many-body theory.  
\end{abstract}


\maketitle

\section{Introduction}
\label{intro}

In a pioneering paper~\cite{pendry1}, Echenique and Pendry investigated the
observability of Rydberg-like electronic states trapped at metal surfaces via
low-energy electron diffraction (LEED) experiments. They discussed the
lifetime broadening of these image-potential-induced surface states (image
states), and reached the important conclusion that they could, in principle,
be resolved for all members of the Rydberg series.

A few years later, Pendry suggested a new experiment~\cite{pendry2}:
measurement of the bremsstrahlung-radiation spectrum from electrons, with
energies no more than a few tens of electron volts, incident on clean
surfaces, thereby turning incident electrons into emitted photons. This {\it
photon}-emission experiment is simply the time reversal of the photoemission
process and was referred to by Pendry as inverse photoemission, or IPE for
short.

Subsequently, Johnson and Smith~\cite{johnson1} pointed out that image states
were potentially observable by angle-resolved IPE; using this technique, Dose
{\it et al.}~\cite{dose1} and Straub and Himpsel~\cite{straub1} reported the
first conclusive experimental evidence for image-potential bound states at the
(100) surfaces of copper and gold. Since then, several observations of image
states have been made using this technique~\cite{ip1,ip2,ip3,ip4,ip5,ip6,ip7},
and also the more recent high-resolution techniques of two-photon
photoemission (2PPE)~\cite{tppe1,tppe2,tppe3} and time-resolved two-photon
photoemission (TR-2PPE)~\cite{tppe4,tppe5,tppe6}. In
2PPE, intense laser radiation is used to populate an unoccupied  state with
the first photon and to photoionize from the intermediate state with the
second photon. In TR-2PPE, the probe pulse which ionizes the intermediate
state is delayed with respect to the pump pulse which populates it, thus
providing a direct measurement of the intermediate-state lifetime.

\begin{figure}
\includegraphics[scale=0.35]{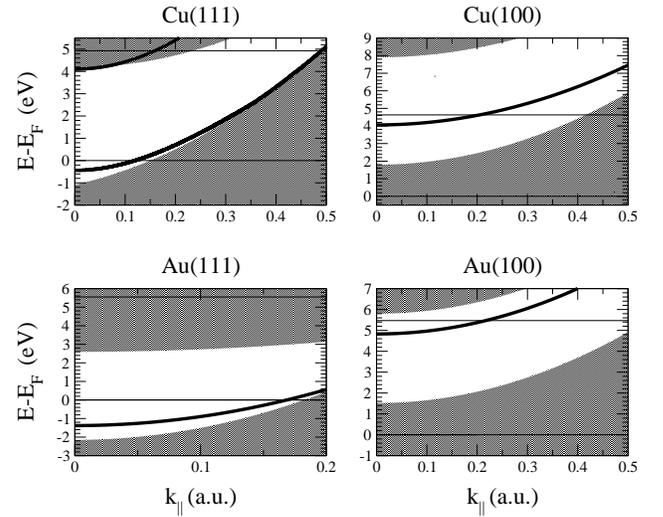}
\caption{The $\Gamma L$ projected bulk band structure (shaded areas) of the
(111) and (100) surfaces of the noble metals Cu and Au. The solid lines
represent Shockley ($n=0$) and image-potential ($n=1$) surface-state bands.
Well-defined Shockley states are only present at the Cu(111) and Au(111) surfaces, where the projected band gap extends below the Fermi level. Well-defined
image states are present at the Cu(100) and Au(100) surfaces, and also at the
(111) surface of Cu. At Au(111) there is only an image-state resonance lying
above the top of the band gap. The horizontal thin solid lines represent Fermi and vacuum levels.}
\label{fig1}
\end{figure}

At metal surfaces, in addition to image states (which are originated in the
combination of the long-range image potential in front of solid surfaces with
the presence of a band gap near the vacuum level)~\cite{image1,image2} there
exist crystal-induced surface states (which would occur even for a step
barrier in the absence of the image potential)~\cite{ss} often classified as
Shockley and Tamm states~\cite{shockley,tamm}: Shockley states exist near the
Fermi level in the projected bulk band gap of simple and
noble metals, and Tamm states exist at the $\bar M$ points of the surface
Brillouin zone for various noble-metal surfaces. The lifetimes of excited holes
at the band edge (${\bf k}_\parallel=0$) of Shockley states have been
investigated with high-resolution angle-resolved photoemission
(ARP)~\cite{arp1,arp2,arp3,arp4} and with the use of the scanning tunneling
microscope (STM)~\cite{stm1,science}. STM techniques have also allowed the
determination of the lifetimes of excited Shockley and image electrons over a
range of energies above the Fermi level~\cite{stm2,stm3}.

Figure~\ref{fig1} illustrates Shockley and image-potential states in the gap of
the $\Gamma L$ projected band structure of the (100) and (111) surfaces of the
noble metals Cu and Au. If an electron or hole is added to the solid at
one of these states, inelastic coupling of the excited quasiparticle with the
crystal may occur through electron-electron (e-e) and electron-phonon (e-ph)
scattering.

In this paper, we first give a brief description of existing calculations of
e-ph inelastic linewidths of image and Shockley states, and we then focus on
the many-body theory of e-e inelastic lifetimes of these states. In
particular, we describe the current status of many-body $GW$ and $GW\Gamma$
calculations, and we report new $GW\Gamma$ calculations of the
self-energy and lifetime of surface states on Au surfaces. We conclude that
short-range exchange-correlation (xc) contributions to the electron (or hole) self-energy are small, as occurs in the case of bulk states. 

Unless otherwise is stated, atomic units are used throughout, i. e.,
$e^2=\hbar=m_e=1$. The atomic unit of length is the Bohr radius,
$a_0=\hbar^2/m_e^2=0.529{\rm\AA}$, the atomic unit of
energy is the Hartree, $1\,{\rm Hartree}=e^2/a_0=27.2\,{\rm eV}$, and
the atomic unit of velocity is the Bohr velocity, $v_0=\alpha\,
c=2.19\times 10^8{\rm cm\,s^{-1}}$, $\alpha$ and $c$ being the fine-structure
constant and the velocity of light, respectively.

\section{Electron-phonon coupling}

The decay rate due to the e-ph interaction, which is relatively important only
in the case of excited Shockley holes near the Fermi level, has been
investigated by using the Eliashberg function~\cite{grimvall}. In particular,
at zero temperature ($T=0$) and in the high-temperature limit
($k_BT>>\omega_m$, $k_B$ being Boltzmann's constant and $\omega_m$, the
maximum phonon frequency) one finds respectively,
assuming translational invariance in the plane of the surface, the following
expressions for the e-ph induced linewidth (or lifetime broadening)
$\Gamma_{ep}$ of surface states of parallel momentum ${\bf k}$ and energy
$E$~\cite{grimvall,eiguren1}: 
\begin{equation}
\Gamma_{ep}^0({\bf k},E)=2\pi\int_0^{{\rm max}(|E|,\omega_m)}\alpha^2F_{\bf k}
(\omega) d\omega
\end{equation}
and
\begin{equation}
\Gamma_{ep}({\bf k},E)=2\pi\,\lambda({\bf k})\,k_BT,
\end{equation}
where $\alpha^2F_{\bf k}(\omega)$ is the Eliashberg function, which represents
a weighted phonon density of states, and
\begin{equation}
\lambda({\bf k})=\int_0^{\omega_m}
{\alpha^2F_{\bf k}(\omega)\over\omega}d\omega.
\end{equation}

For many years, the e-ph contribution $\Gamma_{ep}({\bf k},E)$ to the inelastic
decay of surface states had been calculated using a three-dimensional (3D)
Debye phonon model with $\lambda$ obtained from measurements or calculations of
bulk properties~\cite{grimvall}. More refined calculations, which are based on
an accurate description of the full Eliashberg spectral function, have been
carried out recently by Eiguren {\it et al.}~\cite{eiguren2,eiguren3} for (i)
the Shockley surface-state hole ($n=0$) at the $\bar\Gamma$ point of Al(100)
and the (111) surfaces of the noble metals Cu, Ag, and Au~\cite{eiguren2}, and
(ii) the first ($n=1$) image-state electron at the
$\bar\Gamma$ point of the (100) surfaces of Cu and Ag~\cite{eiguren3}; these
calculations are based on the use of (i) Thomas-Fermi screened Ashcroft
electron-ion pseudopotentials, (ii) single-particle states obtained by solving
a single-particle model one-dimensional (1D) Schr\"odinger equation, and (iii)
a simple force-constant phonon model calculation that yields a phonon spectrum
in good agreement with experimental data.

\begin{table}
\caption{Electron-phonon linewidths at $T=0$ ($\Gamma_{ep}^0$), in meV, and
the parameter $\lambda$ that is responsible for the high-temperature behaviour
of $\Gamma_{ep}$, as reported at the $\bar\Gamma$ point in
Refs.~\cite{eiguren2,eiguren3}.\label{table1}}
\begin{ruledtabular}
\begin{tabular}{llll}
Surface & $n$ & $\Gamma_{ep}^0$ & $\lambda$\\
\hline
Al(100)& 0 & 18 & 0.23\\
Cu(111)& 0 & 7.3& 0.16\\
Ag(111)& 0 & 3.7& 0.12\\
Au(111)& 0 & 3.6& 0.11\\
Cu(100)& 1 & $<1\,{\rm meV}$&$<0.01$\\
Ag(100)& 1 & $<1\,{\rm meV}$&$<0.01$\\
\end{tabular}
\end{ruledtabular}
\end{table}

A summary of the results reported by Eiguren {\it et
al.}~\cite{eiguren2,eiguren3} is presented in Table~\ref{table1}.
Electron-phonon linewidths are particularly relevant in the case of
surface-state holes with energies very near the Fermi level, in which case the
contribution from e-e interactions is very small. In the case of
image states, whose energies lie typically a few electronvolts above the Fermi
level, the e-ph linewidth is found to be $\Gamma_{ep}<1\,{\rm meV}$, thereby
showing the
negligibly small role of phonons in the electron dynamics of image-potential
states.

\section{Electron-electron coupling}

Let us consider an arbitrary many-electron system of density
$n_0({\bf r})$. In the framework of many-body theory, the e-e linewidth (or
decay rate) $\Gamma_{ee}$ of a quasiparticle (electron or hole) that has been 
added in the single-particle state $\psi_i({\bf r})$ of energy 
$\varepsilon_i$ is obtained as the projection of the imaginary part of 
the self-energy $\Sigma({\bf r},{\bf r}';\varepsilon_i)$ over the 
quasiparticle-state itself~\cite{review1,review2}:
\begin{equation}\label{eq1}
\Gamma_{ee}=\mp 2\int d{\bf r}\int d{\bf r}'\psi^{*}_i({\bf r})
{\rm Im}\Sigma({\bf r},{\bf r}';\varepsilon_i)\psi_i({\bf r}'),
\end{equation}
where the $\mp$ sign in front of the integral should be taken to be
minus or plus depending on whether the quasiparticle is an electron
($\varepsilon_i\ge\varepsilon_F$) or a hole
($\varepsilon_i\le\varepsilon_F$), respectively, $\varepsilon_F$ being
the Fermi energy. Alternatively, Eq.~(\ref{eq1}) can be written as follows
\begin{equation}\label{eq2}
\Gamma_{ee}=\mp {2\over\pi}\int_{\varepsilon_i-0^+}^{\varepsilon_i+0^+}
\int d{\bf r}\int d{\bf r}'g^0({\bf r},{\bf r}';\varepsilon)
\,{\rm Im}\Sigma({\bf r},{\bf r}';\varepsilon_i),
\end{equation}
where
\begin{equation}\label{eq3}
g^0({\bf r},{\bf r}';\varepsilon)={i\over 2}\left\{G^0({\bf r}',{\bf r};
\varepsilon)-\left[G^0({\bf r},{\bf r}';\varepsilon)\right]^*\right\},
\end{equation} 
$G^0({\bf r},{\bf r}';\varepsilon)$ being the one-particle Green function of a
noninteracting many-electron system:
\begin{equation}\label{eq4}
G^0({\bf r},{\bf r}';\varepsilon)=\sum_f{\psi_f({\bf r})\psi_f^*({\bf r}')\over
\varepsilon-\varepsilon_f+i\eta}.
\end{equation}
Here, $\psi_f(\bf r)$ and $\varepsilon_f$ represent the complete set of
eigenfunctions and eigenvalues of a one-particle hamiltonian describing the
noninteracting many-electron system. 

\subsection{Self-energy: $G^0W$ and $G^0W^0$ approximations}
 
To lowest order in a series-expansion of the self-energy in terms of 
the frequency-dependent time-ordered screened interaction
$W({\bf r},{\bf r}';\omega)$, the self-energy
$\Sigma({\bf r},{\bf r}';\varepsilon_i)$ is obtained by integrating the product
of the interacting Green function $G({\bf r},{\bf r}';\varepsilon_i-\omega)$
and the screened interaction $W({\bf r},{\bf r}';\omega)$, and is therefore
called the $GW$ self-energy. If one further replaces the interacting Green
function by its noninteracting counterpart
$G^0({\bf r},{\bf r}';\varepsilon_i-\omega)$, one finds the $G^0W$ self-energy.
For the imaginary
part, one can write
\begin{equation}\label{eq5}
{\rm Im}\Sigma^{G^0W}({\bf r},{\bf r}';\varepsilon_i)=\sum_f{'}\,
\psi_{f}^*({\bf r}'){\rm Im}\,W({\bf r},{\bf
r}';|\varepsilon_i-\varepsilon_f|)\,\psi_{f}({\bf r}),
\end{equation}
where the prime in the summation indicates that the sum is extended, as in Eq.~(\ref{eq4}), over a complete set of single-particle states $\psi_f({\bf r})$ of energy
$\varepsilon_f$ but now with the restriction
$\varepsilon_F\leq\varepsilon_f\leq\varepsilon_i$ or
$\varepsilon_i\leq\varepsilon_f\leq \varepsilon_F$. In terms of the
one-particle
noninteracting Green function $G^0({\bf r},{\bf r}';\varepsilon)$, one finds
\begin{eqnarray}\label{eq6}
{\rm Im}\Sigma^{G^0W}({\bf r},{\bf r}';\varepsilon_i)&=&{1\over\pi}
\int_0^{|\varepsilon_i-\varepsilon_F|}d\varepsilon\,
{\rm Im}g^0({\bf r},{\bf r}';|\varepsilon_i-\varepsilon|)\cr\cr
&\times&{\rm Im}\,W({\bf r},{\bf r}';\varepsilon),
\end{eqnarray}
where $g^0({\bf r},{\bf r}';\varepsilon)$ is given by Eq.~(\ref{eq3}).  
Introducing either Eq.~(\ref{eq5}) or Eq.~(\ref{eq6}) into Eq.~(\ref{eq1}) or
Eq.~(\ref{eq2}), one finds an expression for the
e-e linewidth that exactly coincides with the result one would
obtain from the lowest-order probability per unit time for an excited
electron or hole in an initial state $\psi_i({\bf r})$ of energy
$\varepsilon_i$ to be scattered into the state $\psi_f({\bf r})$ of
energy $\varepsilon_f$ by exciting a Fermi system of interacting
electrons from its many-particle ground state to some many-particle
excited state~\cite{pi1}.

The screened interaction $W({\bf r},{\bf r}';\omega)$ entering
Eqs.~(\ref{eq5}) and (\ref{eq6}) can be rigorously expressed as follows
\begin{eqnarray}\label{eq7}
W({\bf r},{\bf
r}';\omega)&=&v({\bf r},{\bf r}')+\int{\rm d}{\bf r}_1\int{\rm d}{\bf
r}_2\,v({\bf r},{\bf r}_1)\cr\cr&\times&\chi({\bf r}_1,{\bf
r}_2;\omega)\,v({\bf r}_2,{\bf r}'),
\end{eqnarray}
$v({\bf r},{\bf r}')$ representing the bare Coulomb interaction and
$\chi({\bf r},{\bf r}';\omega)$ being the time-ordered density-response
function of the many-electron system, which for the positive 
frequencies ($\omega>0$) entering Eqs.~(\ref{eq5}) and (\ref{eq6}) coincides
with the retarded density-response function of linear-response theory. In the
framework of time-dependent density-functional theory (TDDFT)~\cite{tddft},
the {\it exact} retarded density-response function is obtained by solving the
following integral equation~\cite{tddftg}:
\begin{eqnarray}\label{rpa}
&&\chi({\bf r},{\bf r}';\omega)=\chi^0({\bf r},{\bf 
r}';\omega)+\int{\rm d}{\bf
r}_1\int{\rm d}{\bf r}_2\,\chi^0({\bf r},{\bf r}_1;\omega)\cr
\cr&&\times\left\{v({\bf r}_1,{\bf r}_2)+f^{xc}[n_0]({\bf r}_1,{\bf
r}_2;\omega)\right\}\chi({\bf r}_2, {\bf r}';\omega).
\end{eqnarray}
Here, $\chi^0({\bf r},{\bf r}';\omega)$ denotes the density-response
function of noninteracting Kohn-Sham electrons, i.e., independent
electrons moving in the effective Kohn-Sham potential of density-functional
theory (DFT)~\cite{dft}:
\begin{eqnarray}\label{chi0}
\chi^0({\bf r},{\bf r}';\omega)
&=&{2\over\Omega}\sum_{i,j}(f_i-f_j)\cr\cr
&\times&{\psi_i({\bf r})\psi_j^*({\bf r})
\psi_j({\bf r}')\psi_i^*({\bf r}')\over
\omega-\varepsilon_j+\varepsilon_i+{\rm i}\eta},
\end{eqnarray}
where $\Omega$ represents a normalization volume, $f_i$ are Fermi-Dirac
occupation factors [which at zero temperature take the
form $f_i=\Theta(\varepsilon_F-\varepsilon_i)$, $\Theta(x)$ being the Heaviside
step function], and $\psi_i({\bf r})$ and
$\varepsilon_i$ represent the eigenfunctions and eigenvalues of the
Kohn-Sham Hamiltonian of DFT. The other ingredient that is needed
in order to solve Eq.~(\ref{rpa}) is the xc kernel 
$f^{xc}[n_0]({\bf r},{\bf r}';\omega)$, which is the functional derivative of
the {\it unknown} frequency-dependent xc potential $V_{xc}[n]({\bf r},\omega)$
of TDDFT, to be evaluated at $n_0({\bf r})$.

In the random-phase approximation (RPA),
$f^{xc}[n_0]({\bf r},{\bf r}';\omega)$ is set equal to zero:
\begin{eqnarray}\label{rpa2}
&&\chi^{RPA}({\bf r},{\bf r}';\omega)=\chi^0({\bf r},{\bf 
r}';\omega)+\int{\rm d}{\bf
r}_1\int{\rm d}{\bf r}_2\,\chi^0({\bf r},{\bf r}_1;\omega)\cr
\cr&&\times v({\bf r}_1,{\bf r}_2)\,\chi^{RPA}({\bf r}_2, {\bf r}';\omega),
\end{eqnarray}
and the screened interaction $W({\bf r},{\bf r}';\omega)$ is replaced by
\begin{eqnarray}\label{rpa3}
W^0({\bf r},{\bf r}';\omega)&=&v({\bf r},{\bf r}')+\int{\rm d}{\bf r}_1\int{\rm
d}{\bf r}_2\,v({\bf r},{\bf r}_1)\cr\cr&\times&\chi^{RPA}({\bf r}_1,{\bf
r}_2;\omega)\,v({\bf r}_2,{\bf r}'),
\end{eqnarray}
or, equivalently,
\begin{eqnarray}\label{rpa4}
W^0({\bf r},{\bf r}';\omega)&=&v({\bf r},{\bf r}')+\int{\rm d}{\bf r}_1\int{\rm
d}{\bf r}_2\,v({\bf r},{\bf r}_1)\cr\cr&\times&\chi^0({\bf r}_1,{\bf
r}_2;\omega)\,W^0({\bf r}_2,{\bf r}';\omega),
\end{eqnarray}
which yields the so-called  $G^0W^0$ (or $G^0W$-RPA) self-energy:
\begin{eqnarray}\label{eq5p}
{\rm Im}\Sigma^{G^0W^0}({\bf r},{\bf r}';\varepsilon_i)&=&\sum_f{'}\,
\psi_{f}^*({\bf r}'){\rm Im}\,W^0({\bf r},{\bf
r}';|\varepsilon_i-\varepsilon_f|)\cr\cr
&\times&\psi_{f}({\bf r}),
\end{eqnarray}
or, equivalently:
\begin{eqnarray}\label{eq6p}
{\rm Im}\Sigma^{G^0W^0}({\bf r},{\bf r}';\varepsilon_i)&=&{1\over\pi}
\int_0^{|\varepsilon_i-\varepsilon_F|}d\varepsilon\,{\rm Im}
g^0({\bf r},{\bf r}';|\varepsilon_i-\varepsilon|)\cr\cr
&\times&{\rm Im}\,W^0({\bf r},{\bf r}';\varepsilon).
\end{eqnarray}

\subsection{Self-energy: $GW\Gamma$ approach}

The xc kernel $f^{xc}[n_0]({\bf r},{\bf r}';\omega)$ entering Eq.~(\ref{rpa}),
which is absent in the RPA, accounts for the presence of an xc hole associated
to 
all screening electrons in the Fermi sea. Hence, one might be tempted to
conclude that the full $G^0W$ approximation [with the formally exact
screened interaction $W$ of Eq.~(\ref{eq7})] should be a better
approximation than its $G^0W^0$ counterpart (with the screened
interaction $W$ evaluated in the RPA). However, the xc hole associated
to the excited electron (or hole) is still absent in the $G^0W$ approximation.
Therefore, if one goes beyond RPA in the description of
$W$, one should also go beyond the $G^0W$ approximation in the expansion
of the electron self-energy in powers of $W$. By including xc effects
both beyond RPA in the description of $W$ and beyond $G^0W$ in the
description of the self-energy~\cite{mahan,mahan2,delsole1}, the so-called
$GW\Gamma$ approach yields a self-energy that is of the $G^0W$ form:
\begin{equation}\label{eq5pp}
{\rm Im}\Sigma^{GW\Gamma}({\bf r},{\bf r}';\varepsilon_i)=\sum_f{'}\,
\psi_{f}^*({\bf r}'){\rm Im}\,\tilde W({\bf r},{\bf
r}';|\varepsilon_i-\varepsilon_f|)\,\psi_{f}({\bf r}),
\end{equation}
or, equivalently:
\begin{eqnarray}\label{eq6pp}
{\rm Im}\Sigma^{GW\Gamma}({\bf r},{\bf r}';\varepsilon_i)&=&{1\over\pi}
\int_0^{|\varepsilon_i-\varepsilon_F|}d\varepsilon\,{\rm Im}
g^0({\bf r},{\bf r}';|\varepsilon_i-\varepsilon|)\cr\cr
&\times&{\rm Im}\,\tilde W({\bf r},{\bf r}';\varepsilon),
\end{eqnarray}
but with the actual screened interaction $W({\bf r},{\bf r}';\omega)$ entering
Eq.~(\ref{eq5}) being replaced by a new effective screened interaction
\begin{eqnarray}\label{eq:Walda}
&&\tilde W({\bf r},{\bf r}';\omega)
=v({\bf r},{\bf r}')+\int{\rm d}{\bf r}_1\int{\rm
d}{\bf r}_2\,\left\{v({\bf r},{\bf
r}_1)\right.\cr\cr&&+\left.f^{xc}[n_0]({\bf r},{\bf
r}_1;\omega)\right\}\,\chi({\bf r}_1,{\bf r}_2;\omega)\,v({\bf r}_2,{\bf
r}'),
\end{eqnarray}
which includes all powers in $W$ beyond the $G^0W$ approximation.

\subsection{Surface-state wave functions}

\subsubsection{Simple models}

\paragraph{Outside the solid.}

\begin{figure}
\includegraphics[angle=90,scale=0.4]{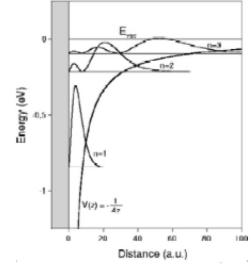}
\caption{Probability amplitudes $\phi_n(z)$ and energies $\varepsilon_n$ of
image-potential-induced bound states ($n=1, 2, 3$) outside an infinitely
repulsive solid surface occupying the $z<0$ space (shaded area). The thick
solid line represents the classical image potential of Eq.~(\ref{classical}).} 
\label{fig2}
\end{figure}

Image states are quantum states trapped in the long-range image-potential well
outside a solid surface that presents a band gap near the vacuum level. In the
case of a metal that occupies the half-space $z<0$, the asymptotic form of the
potential experienced by an electron in the half space $z>0$ is the classical
image potential
\begin{equation}\label{classical}
V(z)=-{1\over 4z}.
\end{equation} 
If one assumes (i) translational invariance in the plane of the surface and
(ii) that due to the presence of a wide band gap at $z<0$ the solid surface is
infinitely repulsive, i.e., $V(z)\to\infty$ at $z<0$~\cite{note1}, then one
easily finds that the solutions of the corresponding one-particle
Schr\"odinger equation
represent a Rydberg-like series of image-potential induced bound states (see
Fig.~\ref{fig2}) of the form:
\begin{equation}\label{hydro}
\psi_{{\bf k},n}({\bf r})={1\over\sqrt A}\,\phi_n(z)\,{\rm e}^
{i{\bf k}\cdot{\bf r}_\parallel},
\end{equation}
with energies
\begin{equation}\label{par}
E_{{\bf k},n}=\varepsilon_n+{\bf k}^2/2,
\end{equation}
where
\begin{equation}\label{hydrogen}
\phi_n(z)\sim z\,\phi_n^{\rm hydrogen}(z/4),\,\,n=1,2,...
\end{equation}
and
\begin{equation}
\varepsilon_n=-{1\over 32n^2},\,\, n=1, 2, ...,
\end{equation}
$\phi_n^{\rm hydrogen}(z)$ representing the well-known wave functions of all
possible $s$-like ($l=0$) bound states of the hydrogen atom.
Here, ${\bf r}_\parallel$ and ${\bf k}$ represent the position and the
wavevector in the $xy$ surface plane.  

\paragraph{Inside the solid.}

\begin{figure}
\includegraphics[scale=0.35]{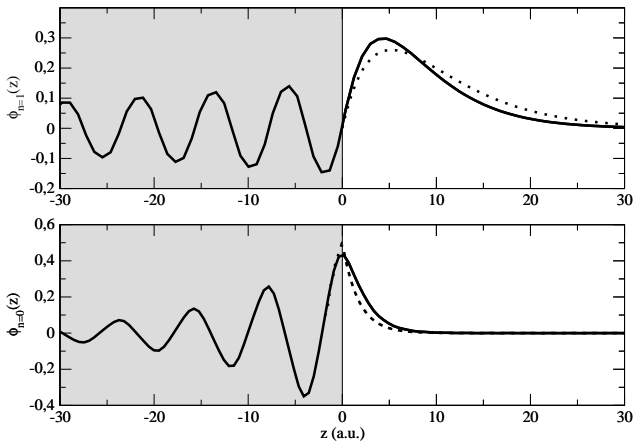}
\caption{Probability amplitudes $\phi_n(z)$ of (a) Shockley ($n=0$) and (b)
image ($n=1$) states on Cu(111). The dotted lines represent the results one
obtains by matching at $z=0$ the wave function of Eqs.~(\ref{nfe}) and
(\ref{nfe2}) to a mere exponential (for the Shockley state) or to the
hydrogenic form of Eq.~(\ref{hydrogen}) with $n=1$ (for the $n=1$ image state). The solid lines
represent the results one obtains by solving a 1D Schr\"odinger equation with
the model potential of Chulkov {\it et al.}~\cite{chulkov1}.} 
\label{fig3}
\end{figure}

In the interior of the solid ($z<0$), both image and Shockley surface states
can be described within a two-band approximation to the nearly-free-electron
(NFE) band structure of the solid~\cite{ashcroft}. Assuming translational
invariance in the plane of
the surface and for a gap that is opened by potential Fourier components
corresponding to reciprocal lattice vectors that are normal to the
surface, surface-state wave functions within the crystal band gap take the form
\begin{equation}\label{nfe}
\phi(z<0)\approx{\rm e}^{\Delta}\,\cos(Gz+\delta).
\end{equation}
Here, $G$ represents the limit of the Brillouin zone in the direction normal to
the surface, and
\begin{equation}\label{nfe2}
\Delta={1\over G}\sqrt{{1\over 4}V_{g}^2-\bar\varepsilon^2},
\end{equation}
where $V_{g}$ and $\bar\varepsilon$ denote the energy gap and the surface-state
energy with respect to the mid gap, respectively, and $\delta$ represents a
phase shift which in the presence of a Shockley-inverted band gap~\cite{forst}
varies from $-\pi/2$ for a surface-state energy $\varepsilon$ at the bottom of
the gap to 0 for a surface-state energy at the top of the gap. Matching at
$z=0$ to a wave function of the hydrogenic-like form of Eq.~(\ref{hydrogen})
(in
the case of image states) or to a mere exponential (in the case of Shockley
states)~\cite{osma}, one finds the wave functions $\phi_n(z)$ plotted by dotted
lines in Fig.~\ref{fig3} for Cu(111).

\subsubsection{One-dimensional model potentials}
\label{sec:chulkov}

Still assuming translational invariance in two directions, i.e., assuming that
the charge density and one-electron potential are constant in the plane of the
surface, Chulkov {\it et al.}~\cite{chulkov1} devised a simplified model that
allows for realistic calculations while retaining at the same time the
essential physics of electron and hole dynamics at solid
surfaces. In the bulk region, this one-dimensional (1D) model
potential is described by a cosine
function which opens the energy gap on the surface of interest, the position
and amplitude of this function being chosen to reproduce the energy gap
observed experimentally and/or obtained from first-principles calculations at
the $\bar\Gamma$ point. At the solid-vacuum interface, it is represented by a
smooth cosine-like function that reproduces the experimental energy of the
Shockley surface state. Finally,
in the vacuum region this 1D potential merges into the long-range classical
image potential of the form of Eq.~(\ref{classical}) in such a way that the
experimental binding energy of the first image state is reproduced. This model
potential has been constructuted for
several metal surfaces~\cite{chulkov2}, and has been used widely for the
investigation of electron and hole dynamics in a variety of situations. 

The $n=0$ and $n=1$ eigenfunctions of a single-particle 1D hamiltonian that
includes the model potential of Chulkov {\it et al.} for Cu(111) are plotted
in Fig.~\ref{fig3} by solid lines, together with the NFE Shockley ($n=0$) and
first image-state ($n=1$) wave functions described in the preceding section.
In the bulk region, these wave functions coincide with the approximate NFE
wave functions (represented in Fig.~\ref{fig3} by dotted lines); however, in
the vacuum region the $n=1$ hydrogenic-like wave function of Eq.~(\ref{hydrogen}) appears to be too little localized near the surface. The
$n=1$ eigenfunction of Chulkov's 1D hamiltonian for Cu(100)~\cite{chulkov0}
was found to reproduce accurately the average probability density derived for
that image state by Hulbert {\it et al.}~\cite{hulbert} from a
first-principles calculation. 

\begin{figure}
\includegraphics[scale=0.35]{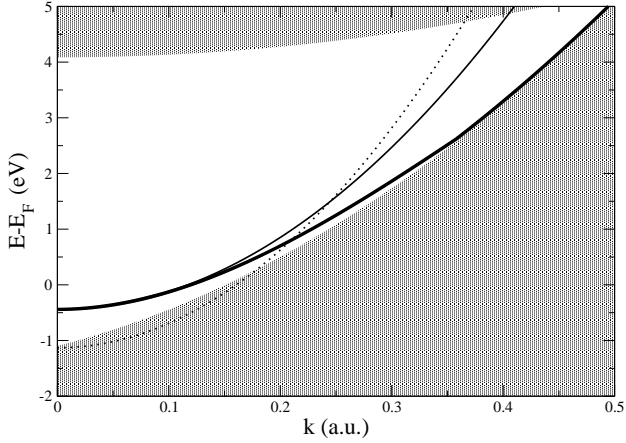}
\caption{Dispersion of the Cu(111) Shockley surface state (thick solid line),
as obtained from 3D {\it ab initio} calculations~\cite{maia1}. Shaded areas
represent areas outside the band gap, where bulk states exist. The thin solid
and dotted lines represent approximate energy dispersions of the Shockley
surface state and the bottom of the projected band gap, respectively, as
obtained from an equation of the form of Eq.~(\ref{par}) with the Cu(111)
$\bar\Gamma$-point Shockley-state energy (with respect to the Fermi level)
$\varepsilon=-0.44\,{\rm eV}$ and the effective mass $m=0.42$ (thin solid
line), and with the Cu(111) $\bar\Gamma$-point bottom-of-the-gap energy (with
respect to the Fermi level) $\varepsilon=-1.09\,{\rm eV}$ and the effective
mass $m=0.22$ (thin dotted line).} 
\label{fig4}
\end{figure}

The assumption that the charge density and one-electron potential are constant
in the plane of the surface is valid for the description of image states,
since their wave functions lie mainly at the vacuum side of the surface and
the electrons move, therefore, in a region with little potential variation
parallel to the surface. Shockley and bulk states, however, do suffer a
significant potential variation in the plane of the surface. In order to
account approximately for this variation, the original 1D model potential of
Chulkov {\it et al.}~\cite{chulkov1}, which had been introduced to describe
the projected band structure at the $\bar\Gamma$ point, was modified along
with the introduction in Eq.~(\ref{par}) of a realistic effective mass for the
dispersion curve of both bulk and surface states~\cite{osma}. Within this
model, however, all Shockley states have the same effective mass, so the
projected band structure is still inaccurate, especially at energies above the
Fermi level, as shown in Fig.~\ref{fig4} for Cu(111).

As an alternative to the 1D model potential of Chulkov {\it et
al.}~\cite{chulkov1}, Vergniory {\it et al.}~\cite{maia1} introduced a
${\bf k}$-dependent 1D potential that is constructed to reproduce the
actual bulk energy bands and surface-state energy dispersion obtained from 3D
first-principles calculations, thereby allowing for a realistic description of
the electronic orbitals beyond the $\bar\Gamma$ point:
\begin{equation}\label{model}
V_k(z)=\cases{U_k+2V_k\cos(2\pi z/a_s),&$z<z_k$\cr\cr
\Phi&$z>z_k$.}
\end{equation} 
Here, $U_k$ and $V_k$ are fitted to the bulk energy bands, $a_s$
represents the interlayer spacing, $\Phi$ is the experimentally
determined work function, and the matching plane $z_k$ is chosen to
give the correct surface-state dispersion.

\begin{figure}
\includegraphics[scale=0.35]{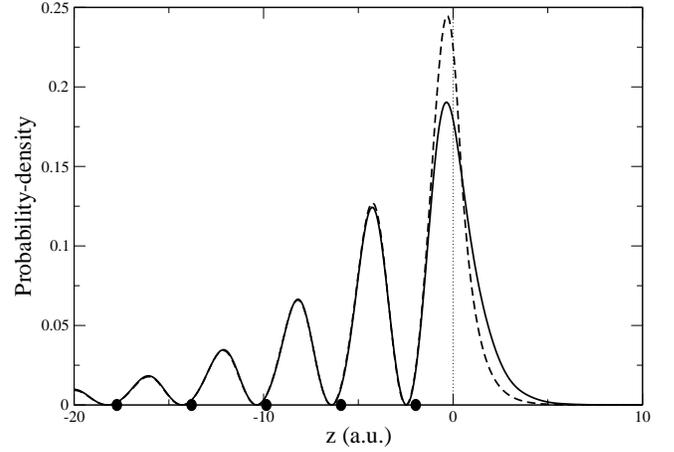}
\caption{Probability-density $|\phi(z)|^2$ of the Shockley surface state at the
center of the surface Brillouin zone ($\bar\Gamma$ point) of Cu(111), as
obtained with the use of two different 1D model potentials: (i) the model
potential of Chulkov {\it et al.}~\cite{chulkov1}, which includes the image
tail outside the surface but fails to reproduce the actual band structure
beyond the
$\bar\Gamma$ point (solid line), and (ii) the model potential of
Eq.~(\ref{model}), which does not include the image tail outside the surface
but is constructed to reproduce the actual bulk energy bands and surface-state
energy dispersion beyond the $\bar\Gamma$ point (dashed line). Full circles represent the atomic positions of Cu in the (111) direction. The geometrical (jellium) electronic edge ($z=0$) has been chosen to be located half an interlayer spacing beyond the last atomic layer.} 
\label{fig5}
\end{figure}

The abrupt 1D step model potential of Eq.~(\ref{model}), which does not account
for the image tail outside the surface, could not possibly be used to describe
image states. However, it has proved to be accurate for the description of
Shockley surface states, which are known to be rather insensitive to the actual
shape of the potential far outside the surface; indeed, the model potential of
Eq.~(\ref{model}) is found to yield a surface-state probability density
$|\phi(z)|^2$ at the band edge ($\bar\Gamma$ point, i.e., ${\bf
k}=0$) of the Shockley surface-state band of Cu(111) that is in
reasonably good agreement with the more realistic surface-state probability
density obtained at $\bar\Gamma$ from the 1D model potential of Chulkov {\it
et al.}~\cite{chulkov1}, as shown in Fig.~\ref{fig5}. Both
probability densities coincide within the bulk, although the probability
density obtained from the step model potential of Eq.~(\ref{model}) appears to
be slightly more localized near the surface, as expected. For the overlap
integral one finds $<\phi_1|\phi_2>=0.99$, $\phi_1$ and $\phi_2$ being the
Shockley probability amplitudes leading to the probability densities
represented in Fig.~\ref{fig5} by solid and dashed lines, respectively. 

\subsection{Screened interaction}

The retarded counterpart of the density-response function entering
Eq.~(\ref{eq7}), which in the framework of TDDFT can be obtained rigorously by
solving the intergral Eq.~(\ref{rpa}), yields, within linear-response theory,
the electron density $\delta n({\bf r},\omega)$ induced in a many-electron
system by a
frequency-dependent external potential $\phi^{ext}({\bf r},\omega)$: 
\begin{equation}\label{eqa1}
\delta n({\bf r},\omega)=\int{\rm d}{\bf r}'\,
\chi({\bf r},{\bf  r'};\omega)\,\phi^{ext}({\bf r'},\omega).
\end{equation}
Hence, the retarded counterpart of the screened interaction
$W({\bf r},{\bf r}';\omega)$ of Eq.~(\ref{eq7}) yields, within linear-response
theory, the total potential $\phi({\bf r},\omega)$ of a unit test charge at
point ${\bf r}$ in the presence of an external test charge of density
$n^{ext}({\bf r},\omega)$:
\begin{equation}\label{tpotnew}
\phi({\bf r},\omega)=\int d{\bf r}'\,W({\bf r},{\bf r}';
\omega)\,n^{ext}({\bf r}',\omega),
\end{equation}
which can also be expressed as follows
\begin{equation}\label{tpot2}
\phi({\bf r},\omega)=\int d{\bf r}'\,\epsilon^{-1}({\bf r},{\bf r}';
\omega)\
\phi^{ext}({\bf r},\omega),
\end{equation}
with
\begin{equation}\label{epsilon}
\epsilon^{-1}({\bf r},{\bf r}';\omega)=\delta({\bf r}-{\bf
r}')+\int{\rm d}{\bf r}''\,
v({\bf r}-{\bf r}'')\,\chi({\bf r}'',{\bf r}';\omega).
\end{equation}
This is the so-called inverse dielectric function of the many-electron system,
whose poles dictate the occurrence of collective electronic excitations.

\subsubsection{Classical model}

In a classical model consisting of a semiinfinite solid at $z<0$ characterized
by a local (frequency-dependent) dielectric function $\epsilon(\omega)$
separated by a planar surface from a semiinfinite vacuum at $z>0$, the total
potential
$\phi({\bf r},\omega)$ at each medium is a solution of Poisson's equation
\begin{equation}
\nabla^2\phi({\bf r},\omega)=-{4\pi\over\epsilon_i(\omega)}\,
n^{ext}({\bf r},\omega),
\end{equation}
$\epsilon_i$ being $\epsilon(\omega)$ or $1$ depending on
whether the point ${\bf r}$ is located in the solid or in the vacuum,
respectively. Hence, the screened interaction
$W({\bf r},{\bf r}';\omega)$ entering
Eq.~(\ref{tpotnew}) is a solution of the following equation:
\begin{equation}\label{poissonl}
\nabla^2 W({\bf r},{\bf r}';\omega)=-{4\pi\over\epsilon_i(\omega)}
\,\delta({\bf r}-{\bf r}').
\end{equation}
Imposing boundary conditions of continuity of the potential and the normal
component of the displacement vector at the interface, one finds
\begin{equation}
W({\bf r},{\bf r}';\omega)=\int d{\bf q}\,
{\rm e}^{i{\bf q}\cdot({\bf r}_\parallel-{\bf r}_\parallel')}\,
W(z,z';q,\omega),
\end{equation}
where
\begin{widetext}
\begin{equation}\label{clas1}
W(z,z';q,\omega)={2\pi\over q}\cases{
\left[{\rm e}^{-q|z-z'|}+g(\omega)\,{\rm e}^{-q(|z|+
z'|)}\right]/\epsilon(\omega),&$z<0,z'<0$,\cr\cr
2\,g(\omega)\,{\rm e}^{-q
z-z'|}/\left[\epsilon(\omega)-1\right],&$z^<<0,z^>>0$,\cr\cr
{\rm e}^{-q|z-z'|}-g(\omega)\,{\rm e}^{-q(|z|+|z'|)},&$z>0,z'>0$,}
\end{equation}
\end{widetext}
$z^<$ ($z^>$) being the smallest (largest) of $z$ and $z'$, and $g$
being the classical surface-response function:
\begin{equation}\label{g1}
g(\omega)={\epsilon(\omega)-1\over
\epsilon(\omega)+1}.
\end{equation} 

An inspection of Eqs.~(\ref{clas1}) and (\ref{g1}) shows that the
screened interaction $W(z,z';q,\omega)$ has poles at the {\it classical}
bulk- and surface-plasmon conditions dictated by $\epsilon(\omega)=0$ and by
$\epsilon(\omega)+1=0$, respectively~\cite{plasmons}. Since e-e inelastic linewidths of
Shockley and image states are typically dominated by the excitation of
electron-hole (e-h) pairs and not by the excitation of plasmons (whose energies
are typically too large)~\cite{notearan}, the classical
screened interaction of Eq.~(\ref{clas1}) (which obviously does not account for
the excitation of e-h pairs) is of no use in this context.

\subsubsection{Specular-reflection model (SRM)}

A simple scheme that gives account of the excitation of e-h pairs, and has the
virtue of expressing the screened interaction $W(z,z';q,\omega)$ in terms of
the dielectric function $\epsilon(q,\omega)$ of a homogeneous electron gas
representing the bulk material, is the so-called specular-reflection model
reported independently by Wagner~\cite{wagner} and by Ritchie and
Marusak~\cite{marusak}. In this model, the semi-infinite solid is described by
an electron gas in which all electrons
are considered to be specularly reflected at the surface, thereby the electron
density vanishing outside. One finds:
\begin{widetext}
\begin{equation}\label{gh1}
W(z,z';q,\omega)={2\pi\over q}\cases{
\epsilon_s(z-z')+\epsilon_s(z+z')-2\,g(q,\omega)\,
\displaystyle{\epsilon_s(z)\,\epsilon_s(z')\over 1-\epsilon_s^0},
&$z<0,z'<0$,\cr\cr
2\,g(q,\omega)\,\displaystyle{\epsilon_s(z^<)\over 1-\epsilon_s^0}
\,{\rm e}^{-qz^>},&$z^<<0,z^>>0$,\cr\cr
{\rm e}^{-q|z-z'|}-g(q,\omega)\,
{\rm e}^{-q(z+z')},&$z>0,z'>0$,}
\end{equation}
\end{widetext}
where the surface response function is now given by the following expression:
\begin{equation}\label{gsrm}
g(q,\omega)={1-\epsilon_s^0(q,\omega)\over 1+\epsilon_s^0(q,\omega)},
\end{equation}
with
\begin{equation}\label{epsilons}
\epsilon_s(z;q,\omega)={q\over\pi}\int_{-\infty}^{+\infty}
{dq_z\over Q^2}\,{\rm e}^{iq_zz}\epsilon^{-1}(Q,\omega),
\end{equation} 
\begin{equation}\label{gh5}
\epsilon_s^0(q,\omega)=\epsilon_s(z=0;q,\omega),
\end{equation}
and $Q=\sqrt{q^2+q_z^2}$. If the $Q$-dependence of the actual
$\epsilon(Q,\omega)$ dielectric function of
a homogeneous electron gas is ignored, the SRM screened interaction of
Eq.~(\ref{gh1}) reduces to the {\it classical} screened interaction of
Eq.~(\ref{clas1}).

The inverse dielectric function $\epsilon^{-1}(Q,\omega)$ entering
Eq.~(\ref{epsilons}) represents the 3D Fourier transform of the
inverse dielectric function $\epsilon^{-1}({\bf r},{\bf r}',\omega)$
of a homogeneous electron gas. From Eq.~(\ref{epsilon}), one finds:
\begin{equation}\label{epsilonq}
\epsilon^{-1}(Q,\omega)=1+v_Q\,\chi(Q,\omega),
\end{equation}
where $\chi(Q,\omega)$ represents the 3D Fourier transform of the
density-response function $\chi({\bf r},{\bf r}';\omega)$, and
$v_Q$ is the 3D Fourier transform of the bare Coulomb interaction
$v({\bf r},{\bf r}')$: $v_Q=4\pi/Q^2$.

In the framework of TDDFT, one uses Eq.~(\ref{rpa}) to find
\begin{eqnarray}\label{chiq}
&&\chi(Q,\omega)=\chi^0(Q,\omega)+
\chi^0(Q,\omega)\cr\cr
&&\times\left\{v_Q+f^{xc}(n_0;Q,\omega)\right\}\
\chi(Q,\omega),
\end{eqnarray}
with $\chi^0(Q,\omega)$ and $f^{xc}(n_0;Q,\omega)$ being the 3D
Fourier transforms of the noninteracting density-response function
$\chi^0({\bf r},{\bf r}';\omega)$ and the xc kernel
$f^{xc}[n_0]({\bf r},{\bf r}';\omega)$ of a homogeneous electron gas of density
$n_0$. For a homogeneous electron gas, the eigenfunctions
$\psi_i({\bf r})$ entering Eq.~(\ref{chi0}) are all plane waves; thus,
the integrations can be carried out analytically to yield
the well-known Lindhard function
$\chi^0(Q,\omega)$~\cite{lindhard}.
If one sets the xc kernel $f^{xc}(n_0;Q,\omega)$ equal to zero,
the introduction of
Eq.~(\ref{chiq}) into Eq.~(\ref{epsilonq}) yields the RPA dielectric
function
\begin{equation}\label{chiqrpa}
\epsilon^{RPA}(Q,\omega)=1-v_Q\,\chi^0(Q,\omega),
\end{equation}
which is easy to evaluate.

\subsubsection{1D self-consistent scheme}

For an accurate quantal description of the electronic excitations that occur in
a semi-infinite solid, we need to consider the true self-consistent
density-response function $\chi({\bf r},{\bf r}';\omega)$ entering
Eqs.~(\ref{eq7}) and (\ref{eq:Walda}).

Assuming translational invariance in the plane of the surface, one can still
define the 2D Fourier transforms $W(z,z';q,\omega)$ and
$\tilde W(z,z';q,\omega)$,
which according to Eqs.~(\ref{eq7}) and (\ref{eq:Walda}) can be obtained as
follows
\begin{eqnarray}\label{wsurf}
&&W(z,z';q,\omega)=v(z,z';q)+\int dz_1\int dz_2\,v(z,z_1;q)\cr\cr
&&\times\chi(z_1,z_2;q,\omega)\,v(z_2,z';q),
\end{eqnarray}
and 
\begin{eqnarray}\label{wsurf2}
&&\tilde W(z,z';q,\omega)
=v(z,z';q)+\int{\rm d}z_1\int{\rm d}z_2\,
\left\{v(z,z_1;q)\right.\cr\cr&&
+\left.f^{xc}[n_0](z,z_1;q,\omega)\right\}\,\chi(z_1,z_2;q,\omega)\,
v(z_2,z',q),
\end{eqnarray}
where $v(z,z';q)$ is the 2D Fourier transform of the bare Coulomb
interaction $v({\bf r},{\bf r}')$:
\begin{equation}\label{bare}
v(z,z';q)={2\pi\over q}\,{\rm e}^{-q|z-z'|},
\end{equation}
$f^{xc}[n](z,z';q,\omega)$ is the 2D Fourier transform of the xc kernel
$f^{xc}[n]({\bf r},{\bf r}';\omega)$, and $\chi(z,z';q,\omega)$ denotes the
2D Fourier transform of the interacting density-response function
$\chi({\bf r},{\bf r}';\omega)$. In the framework of TDDFT, one uses
Eq.~(\ref{rpa}) to find:
\begin{eqnarray}\label{eq:Xalda3}
&&\chi(z,z';q,\omega)=\chi^0(z,z';q,\omega)+
\int dz_1\int dz_2\,\chi^0(z,z_1;q,\omega)\cr\cr
&&\times\left\{v(z_1,z_2;q)+f^{xc}[n_0](z_1,z_2;q,\omega)\right\}
\chi(z_2,z';q,\omega),
\end{eqnarray}
where $\chi^0(z,z';q,\omega)$ denotes the 2D Fourier transform of the noninteracting density-response function $\chi^0({\bf r},{\bf r}';\omega)$. Using
Eq.~(\ref{chi0}), and noting that the single-particle orbitals
$\psi_i({\bf r})$ now take the form
\begin{equation}\label{psi0}
\psi_{{\bf k},i}({\bf r})=\phi_i(z)\,{\rm e}^{i{\bf k}\cdot{\bf r}_\parallel},
\end{equation}
one finds:
\begin{eqnarray}
&&\chi^{0}(z,z';q,\omega)={2\over A}\mathrel
{\mathop{\sum}\limits_{i,j}}\phi_{i}(z)\phi_{j}^*(z)\phi_{j}(z')
\phi_{i}^*(z')\cr\cr &&\times \mathop{\sum}\limits_{{\bf k}}
\frac{f_{{\bf k},i}-f_{{\bf k}+{\bf q},j}}{E_{{\bf k},i}- E_{{\bf
k}+{\bf q},j}+\omega+i\eta}, \label{chi0zz1}
\end{eqnarray}
where $f_{{\bf k},i}$ are Fermi-Dirac occupation factors [which at zero
temperature take the form $f_{{\bf k},i}=\Theta(\varepsilon_F-E_{{\bf k},i})$], and
\begin{equation}\label{e0}
E_{{\bf k},i}=\varepsilon_i+{k^2\over 2},
\end{equation}
the single-particle orbitals $\phi_i(z)$ and energies $\varepsilon_i$
now being the eigenfunctions and eigenenergies of a 1D Kohn-Sham hamiltonian.
In order to account for the actual band structure of $sp$ electrons near the
surface of simple and noble metals, $\phi_i(z)$ and $\varepsilon_i$ have been
succesfully taken to be the solutions of the 1D single-particle Schr\"odinger
equation of Chulkov {\it et al.}~\cite{chulkov1} described in the previous
section.

\subsubsection{Asymptotics}

For $z$ and $z'$ coordinates that are far from the surface into the vacuum,
where the electron density vanishes, Eq.~(\ref{wsurf}) takes the
form~\cite{notesrm} 
\begin{equation}\label{wg}
W(z,z';q,\omega)=v(z,z';q)-{2\pi\over q}\,{\rm e}^{-q(z+z')}\,
g(q,\omega),
\end{equation}
with the surface-response function $g(q,\omega)$ now being given by the general
expression~\cite{liebsch0}:
\begin{equation}\label{g}
g(q,\omega)=-{2\pi\over q}\,\int dz_1\int dz_2\,{\rm e}^{q(z_1+z_2)}
\,\chi(z_1,z_2;q,\omega).
\end{equation}

Persson and Anderson~\cite{pa} and Persson and Zaremba~\cite{pz} investigated
the structure of the so-called surface loss function ${\rm Im}g(q,\omega)$ for
small $q$ and $\omega$. Persson and Zaremba found the following approximate
result~\cite{pz}:
\begin{equation}
{\rm Im}\,g=({\rm Im}\,g)_{surf}+({\rm Im}\,g)_{bulk}+({\rm Im}\,g)_{int},
\end{equation}
where $({\rm Im}\,g)_{surf}$ and $({\rm Im}\,g)_{bulk}$ represent contributions
from surface and bulk excitation of e-h pairs, and $({\rm Im}\,g)_{int}$
represents the contribution to the surface loss function coming from the
interference
between the bulk and surface excitations:
\begin{equation}\label{gsurface}
({\rm Im}\,g)_{surf}=2\xi{q\over k_F}{\omega\over\omega_p},
\end{equation}
\begin{equation}\label{gbulk}
({\rm Im}\,g)_{bulk}={1\over 2}
m_{opt}^2\left({\omega\over\omega_p}\right)^2\eta^3G(\eta),
\end{equation}
and
\begin{equation}\label{ginterference}
({\rm Im}\,g)_{int}=-{8\over\pi^2}{m_{opt}\over k_F}
\left({\omega\over\omega_p}\right)^2\eta{h\over 1+\eta^2},
\end{equation}
with $\eta=\omega/(qk_F)$ and
\begin{equation}\label{geta}
G(\eta)=8\cases{1,
&for $\eta<1$,\cr\cr
1-(1+{1\over 2}\eta^{-2})(1-\eta^{-2})^{1/2},&for $\eta>1$.\cr\cr}
\end{equation}
Here, $k_F$ and $\omega_p$ represent the Fermi momentum and the plasmon frequency, respectively: $k_F^3=3\pi^2\bar n_0$ and $\omega_p^2=4\pi\bar n_0$, $\bar n_0$ being the mean electron density. The values of $\xi$,
$m_{opt}$, and $h$ are given in Ref.~\cite{pz}. The surface contribution of
Eq.~(\ref{gsurface}) had already been reported in Ref.~\cite{pa}, the bulk
contribution of Eq.~(\ref{gbulk}) differs from that used in Ref.~\cite{pa} by
the factor of the optical mass and a factor of $1\over 2$ which had been
missed previously, and the contribution $({\rm Im}\,g)_{int}$ had been
neglected in Ref.~\cite{pa}.

\subsubsection{The xc kernel $f^{xc}[n](z,z';q,\omega)$} 
\label{fxc}

Several approximations can be used to evaluate the {\it unknown} xc kernel
$f^{xc}[n_0](z,z';q,\omega)$ entering Eqs.~(\ref{wsurf2}) and
(\ref{eq:Xalda3}).

\paragraph{Random-phase approximation (RPA).}
Nowadays one usually refers to the RPA as the result of simply setting the xc
kernel $f^{xc}[n](z,z';q,\omega)$ equal to zero: $f^{xc}[n](z,z';q,\omega)=0$,
but still using in Eqs.~(\ref{psi0}) and (\ref{e0}) single-particle Kohn-Sham
states and energies $\phi_i(z)$ and $\varepsilon_i$ that go beyond the Hartree
approximation.

\paragraph{Adiabatic local-density approximation (ALDA).}
If one assumes that dynamic electron-density fluctuations are slowly 
varying in all directions, the xc kernel $f^{xc}[n](z,z';q,\omega)$ is
easily found to be given by the following expression~\cite{liebsch0}:
\begin{equation}\label{alda}
f^{xc}[n](z,z';q,\omega)=\bar 
f^{xc}(n=n(z);Q=0,\omega=0)\,\delta(z-z').
\end{equation}
Here, $\bar f^{xc}(n=n(z);Q,\omega)$ is the 3D Fourier transform of 
the xc kernel of a homogeneous electron gas of density $n$ equal to the local
density $n(z)$, which in the limit as $Q\to 0$ and $\omega\to 0$ is known
to be the second derivative of the xc energy per particle
$\varepsilon_{xc}(n)$. One
typically uses parametrizations~\cite{pw} of the diffusion Monte Carlo (DMC)
xc energy $\varepsilon_{xc}$ reported by Ceperley and Alder~\cite{ca}.

\paragraph{Adiabatic {\it nonlocal} approximation (ANLDA).}
The investigation of short-range xc effects in solids has focused to a great
extent onto the homogeneous electron gas~\cite{constantin}. Hence, assuming
that the {\it unperturbed} density variation $\left[n(z)-n(z')\right]$ is
small
within the short range of $f^{xc}[n](z,z';q,\omega)$, one can adopt the
following average adiabatic {\it nonlocal} approximation~\cite{delsole3,lein,pp}: 
\begin{equation}\label{one}
f^{xc}[n](z,z';q,\omega)=\bar 
f^{xc}(\left[n(z)+n(z')\right]/2;z,z';q,\omega=0),
\end{equation}  
where $\bar f^{xc}(n;z,z';q,\omega)$ represents the 1D Fourier transform of the
xc kernel $\bar f^{xc}(n;Q,\omega)$ of a homogeneous electron gas of density
$n$. A parametrization of the accurate DMC calculations reported by Moroni {\it
et al.}~\cite{moroni} for the static ($\omega=0$) $Q$-dependent {\it
nonlocal} xc kernel $\bar f^{xc}(n;Q,\omega=0)$ that satisfies the well-known
small and large-wavelength asymptotic behaviour was carried out by Corradini
{\it et al.} (CDOP)~\cite{corradini}. An explicit expression for the 2D
Fourier transform of the CDOP
parametrization of $\bar f^{xc}(n;Q,\omega=0)$ was reported in
Ref.~\cite{pp}:
\begin{widetext}
\begin{equation}\label{corradini}
\bar f^{xc}(n;z,z';k)=-\frac{4\pi e^2 C}{k_F^2}\delta(\tilde 
z)-\frac{2\pi
e^2 B}{\sqrt{gk_F^2+k^2}}\,{\rm e}^{-\sqrt{gk_F^2+k^2}|\tilde z|}-
\frac{2\alpha\sqrt{\pi/\beta}e^2}{k_F^3}\left[\frac{2\beta-k_F^2\tilde
z^2}{4\beta^2}k_F^2+k^2\right] {\rm e}^{-\beta\left[k_F^2\tilde
z^2/4\beta^2
+k^2/k_F^2\right]},
\end{equation}
\end{widetext}
where $C$, $B$, $g$, $\alpha$, and $\beta$ are dimensionless functions
of the electron density (see Ref.~\cite{corradini}),
$k_F=(3\pi^2n)^{1/3}$, and $\tilde z=z-z'$.

Calculations of the frequency dependence of the xc kernel of a homogeneous
electron gas have been carried out mainly in the limit of long
wavelengths~\cite{v1,v2,v3,v4,v5,v6}, but work has also been done for finite wave vectors~\cite{z1,z2,z3,z4}. Approximate expressions for the
frequency-dependent xc kernel of inhomogeneous systems have been reported in Refs.~\cite{petersilka,burke,delsole2,nazarov}. 

\subsubsection{$d-$electron screening}
\label{delectrons}

The 1D self-consistent scheme described above has proved to be appropriate for
the description of the screened interaction of $sp$ electrons in simple and
noble metals. It has been argued, however, in the past that a realistic
first-principles description of the electronic band structure is of key
importance in the determination of the inelastic lifetime of bulk electronic
states in the noble metals, due to the participation of $d$ electrons in the
screening of e-e interactions~\cite{campillo}.

Following the scheme originally developed by Liebsch to describe the anomalous
dispersion of surface plasmons in Ag~\cite{liebsch1}, Garc\'\i a-Lekue {\it et
al.}~\cite{aran} accounted for the presence of occupied $d$-bands in the noble
metals by assuming that $sp$ valence electrons are embedded in a polarizable
background at $z\le z_0$ characterized by a local dielectric function
$\epsilon_d(\omega)$. Within this model, the bare Coulomb interaction
$v(z,z';q)$ entering Eq.~(\ref{eq:Xalda3}) is replaced by a modified
($d$-screened) Coulomb interaction $v'(z,z';q,\omega)$ whose 2D Fourier
transform yields~\cite{catalina}
\begin{eqnarray}\label{ll} v'(z,z';q,\omega)&=&\frac{2\pi}
{q\,\epsilon_d(z',\omega)}\, [{\rm e}^{-q\,|z-z'|}+{\rm
sgn}(z_d-z')\nonumber\\ &\times&\sigma_d(\omega)\,
{\rm e}^{-q|z-z_d|}{\rm e}^{-q|z_d-z'|}],
\end{eqnarray}
where
\begin{equation}\label{epsd}
\epsilon_d(z,\omega)=\cases{\epsilon_d(\omega),&$z\leq z_d$ \cr\cr 1, & $z >
z_d$}
\end{equation}
and
\begin{equation}
\sigma_d={\epsilon_d(\omega)-1\over\epsilon_d(\omega)+1}.
\end{equation}
The first term in Eq. (\ref{ll}) is simply the 2D Fourier transform of the bare
Coulomb interaction [see Eq.~(\ref{bare})], but now screened by the
polarization charges induced within the polarizable background. The second
term stems from polarization charges at the boundary of the medium. 

\subsubsection{Periodic surface}
\label{periodic}

For a real periodic surface, one may introduce the following Fourier expansion
of
the screened interaction:
\begin{eqnarray}
W({\bf r},{\bf r}';\omega)&=&{1\over A}\sum_{\bf q}^{SBZ}\sum_{{\bf
g},{\bf g}'}
{\rm e}^{i({\bf q}+{\bf g})\cdot{\bf r}_\parallel}
{\rm e}^{-i({\bf q}+{\bf g}')\cdot{\bf r}_\parallel'}\cr\cr
&\times&W_{{\bf g},{\bf g}'}(z,z';{\bf q},\omega),
\end{eqnarray}
where ${\bf q}$ is a 2D wave vector in the surface Brillouin zone
(SBZ),
and ${\bf g}$ and ${\bf g}'$ denote 2D reciprocal-lattice vectors.
According to Eq.~(\ref{eq7}), the 2D Fourier coefficients
$W_{{\bf g},{\bf g}'}(z,z';{\bf q},\omega)$ are given by the following
expression:
\begin{eqnarray}\label{screenedg}
&&W_{{\bf g},{\bf g}'}(z,z';{\bf q},\omega)=
v_{\bf g}(z,z';{\bf q})\,\delta_{{\bf g},{\bf g}'}+
\int dz_1\int dz_2\cr\cr
&&\times v_{\bf g}(z,z_1;{\bf q})\,
\chi_{{\bf g},{\bf g}'}(z_1,z_2;{\bf q},\omega)\,
v_{{\bf g}'}(z_2,z';{\bf q}),
\end{eqnarray}
where $v_{\bf g}(z,z';{\bf q})$ denote
the 2D Fourier coefficients of the bare Coulomb interaction
$v({\bf r},{\bf r}')$:
\begin{equation}
v_{\bf g}(z,z';{\bf q})={2\pi\over|{\bf q}+{\bf g}|}\,
{\rm e}^{-|{\bf q}+{\bf g}|\,|z-z'|},
\end{equation}
and $\chi_{{\bf g},{\bf g}'}(z,z';{\bf q},\omega)$ are the Fourier
coefficients of the interacting density-response function
$\chi({\bf r},{\bf r}';\omega)$. In the framework of TDDFT, one uses
Eq.~(\ref{rpa}) to find:
\begin{eqnarray}\label{eq:Xalda4}
&&\chi_{{\bf g},{\bf g}'}(z,z';{\bf q},\omega)=
\chi_{{\bf g},{\bf g}'}^0(z,z';{\bf q},\omega)+
\int dz_1\int dz_2\cr\cr
&&\times\chi_{{\bf g},{\bf g}'}^0(z,z_1;{\bf q},\omega)
\,\times\left[v_{{\bf g}_1}(z_1,z_2;{\bf q})\,\delta_{{\bf g}_1,{\bf
g}_2}\right.\cr\cr
&&\left.+f_{{\bf g}_1,{\bf g}_2}^{xc}[n_0]
(z_1,z_2;{\bf q},\omega)\right]
\chi_{{\bf g}_2,{\bf g}'}(z_2,z';{\bf q},\omega),
\end{eqnarray}
where $\chi_{{\bf g},{\bf g}'}^0(z,z';{\bf q},\omega)$ and
$f_{{\bf g},{\bf g}'}^{xc}[n_0](z,z';{\bf q},\omega)$ denote the
Fourier coefficients of the noninteracting density-response function
$\chi^0({\bf r},{\bf r}';\omega)$ and the xc kernel
$f^{xc}[n_0]({\bf r},{\bf r}';\omega)$, respectively. Using
Eq.~(\ref{chi0}), one finds:
\begin{eqnarray}\label{eq9}
&&\chi_{{\bf g},{\bf g}'}^0(z,z';{\bf q},\omega)=\frac{2}{A}
\,\sum_{\bf k}^{\rm SBZ}\sum_{n,n'}\frac{f_{{\bf k},n}-
f_{{\bf k}+{\bf q},n'}}
{\varepsilon_{{\bf k},n}-\varepsilon_{{\bf k}+{\bf q},n'} +\omega +
{ i}\eta}\cr\cr
&&\times\left\langle\phi_{{\bf k},n}(z)|e^{-{ i}({\bf q}+{\bf g})\cdot{\bf
r}_\parallel}|\phi_{{\bf k}+{\bf q},n'}(z')\right\rangle\cr\cr
&&\times\left\langle\phi_{{\bf k}+{\bf q},n'}(z')|e^{{ i}({\bf q}+{\bf
g}')\cdot{\bf
r}_\parallel}|\phi_{{\bf k},n}(z)\right\rangle,
\end{eqnarray}
the single-particle orbitals
$\phi_{{\bf k},n}({\bf r})$ and energies
$\varepsilon_{{\bf k},n}$ being the eigenfunctions and eigenvalues
of a 3D Kohn-Sham Hamiltonian with an effective potential that is
periodic in the plane of the surface.

\section{Results and discussion}

\subsection{Image states}

The first quantitative evaluation of image-state lifetimes was reported in
Ref.~\cite{sols}. This calculation was carried out from Eqs.~(\ref{eq1}) and
(\ref{eq5}), with (i) the hydrogenic-like image-state wave function $\phi_i(z)$
of Eq.~(\ref{hydrogen}) with $n=1$ and no penetration into the solid, (ii) the
bulk final state wave functions $\phi_f(z)$ obtained with the use of a step
model potential, and (iii) two simplified models for the screened interaction:
the SRM of Eq.~(\ref{gh1}) with the RPA for the bulk dielectric function, and
the surface response function reported by Persson and Anderson~\cite{pa}. In
subsequent calculations the penetration of the image-state wave function into
the crystal was allowed~\cite{deandres}, and the role that the unoccupied part
of the narrow Shockley surface state on the (111) surfaces of Cu and Ni plays
in the decay of the $n=1$ image state on these surfaces was investigated by
Gao and Lundqvist~\cite{gao}. In this work, the image-state
wave functions were also approximated by hydrogenic-like wave functions of the
form of Eq.~(\ref{hydrogen}) with no penetration into the solid, a simplified
parametrized form was used for the Shockley surface-state wave function, and
screening effects were neglected altogether. A $G^0W^0$ calculation of the imaginary
part of the electron self-energy near a jellium surface was also
reported~\cite{deisz}, showing the key role that a full evaluation of this
quantity may play in the description of surface-state lifetimes.

The first self-consistent many-body calculations of image-state lifetimes on
noble and simple metals were reported by Chulkov {\it et
al.}~\cite{chulkov0,chulkov3}, and good
agreement with the experimentally determined decay
times~\cite{hofer,knoesel,shumay} was found. In
these calculations, all wave functions and energies were obtained by solving a
single-particle Schr\"odinger equation with the physically motivated 1D model
potential of Ref.~\cite{chulkov1}, and the
electron self-energy was evaluated in the $G^0W^0$ approximation. The potential
variation in the plane of the surface was considered later through the
introduction of an effective mass~\cite{osma}, and self-consistent
calculations of the key role that the partially occupied
Shockley surface state plays in the decay of image states on Cu(111) were
carried out~\cite{osma}. The inclusion of xc effects was investigated in
the framework of the $GW\Gamma$ approximation, first with an adiabatic
local-density description~\cite{sarria} and more recently with an adiabatic
{\it non-local} description of the xc kernel~\cite{maia2}.

\begin{figure}
\includegraphics[scale=0.35]{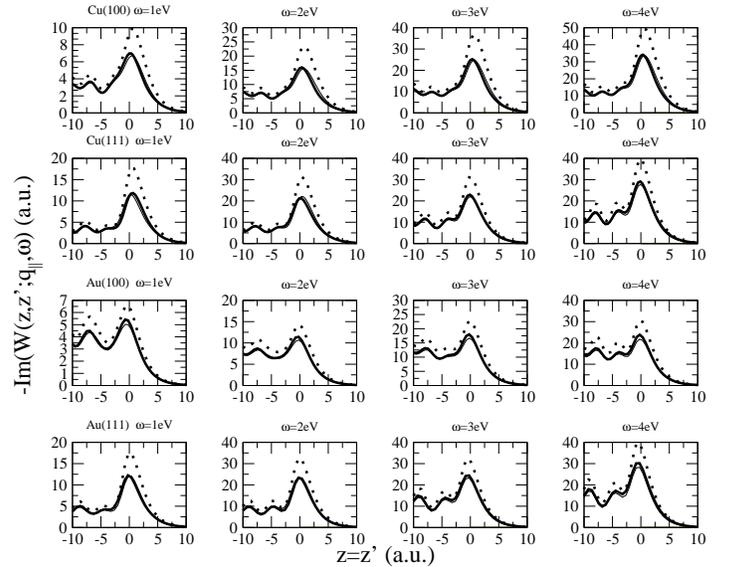}
\caption{Imaginary part of the screened interaction $W(z,z';q,\omega)$
and the effective screened interaction $\tilde W(z,z';q,\omega)$, as a
function of $z=z'$ for a fixed value of $k$ ($q=0.5\,{\rm\AA}^{-1}$) and
various values of $\omega$ ($\omega=1, 2, 3, 4\,{\rm eV}$), in the vicinity of
the (100) and (111) surfaces of Cu and Au. ANLDA calculations of
${\rm Im}\left[\tilde W(z,z';q,\omega)\right]$ are
represented by thick solid lines. RPA and ANLDA calculations of
${\rm Im}\left[W(z,z';q,\omega)\right]$ are represented by thin solid 
and dotted lines, respectively. The RPA ${\rm Im}\left[W(z,z';q,\omega)\right]$
and the ANLDA ${\rm Im}\left[W(z,z';q,\omega)\right]$ are nearly indistinguishable, as a result of a cancellation between the xc effects due to the presence of (i) an xc hole associated with all electrons in the Fermi sea and (ii) an xc hole associated with the excited electron or hole.}
\label{fig6}
\end{figure}

The impact of xc effects on the imaginary part of the effective screened
interaction of Eq.~(\ref{wsurf2}) in the vicinity of the (100) and (111)
surfaces of Cu and Au is illustrated in Fig.~\ref{fig6}, where ANLDA
calculations of ${\rm Im}[\tilde W(z,z';q,\omega)]$ (with full inclusion
of xc effects) are compared to calculations of
${\rm Im}[W(z,z';k,\omega)]$ with (ANLDA)
and without (RPA) xc effects. Exchange-correlation effects included in the
effective screened interaction have two sources. First, there is the reduction
of the screening due to the presence of an xc hole associated with all electrons
in the Fermi sea [see Eq.~(\ref{eq:Xalda3})], which is included in the
calculations represented in Fig.~\ref{fig6} by thick solid lines and also in
the calculations represented by dotted lines. Secondly, there is the reduction
of the effective screened interaction itself due to the xc hole associated with the excited electron or hole [see Eq.~(\ref{wsurf2})], which is only included in the
calculations represented in Fig.~\ref{fig6} by thick solid lines. These
contributions have opposite signs, thereby bringing the ANLDA
${\rm Im}[\tilde W(z,z';k,\omega)]$ (thick solid lines) back to the RPA
${\rm Im}[W(z,z';k,\omega)]$ (thin solid lines).    

\begin{figure}
\includegraphics[scale=0.35]{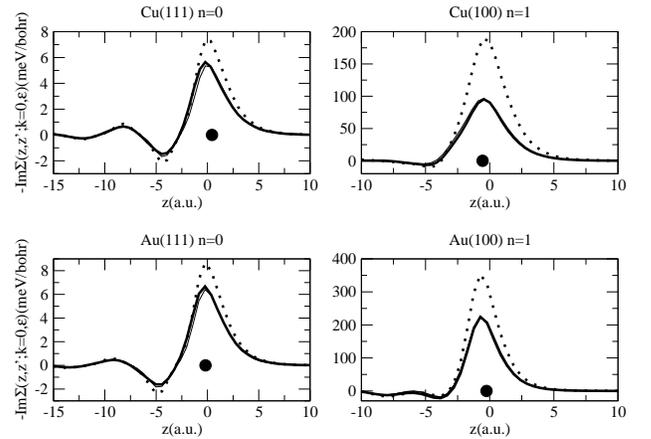}
\caption{$G^0W^0$ ($G^0W$-RPA), $G^0W$, and $GW\Gamma$ calculations of the
imaginary part of the $n=1$ image-state self-energy 
$\Sigma(z,z';{\bf k}=0,\varepsilon_{\bf k})$, versus $z$, in the vicinity of
the (100) surfaces of Cu and Au. The solid circle represents the value of $z'$
in each case. $GW\Gamma$ calculations (as obtained with the use of our ANLDA xc
kernel) are represented by thick solid lines. $G^0W$ (also using our ANLDA xc
kernel) and $G^0W^0$ calculations are represented by dotted and thin solid
lines, respectively. ALDA calculations, not plotted in this figure, are found
to nearly concide with ANLDA calculations.}
\label{fig7}
\end{figure}

Figure~\ref{fig7} exhibits $G^0W^0$, $G^0W$, and $GW\Gamma$ calculations of the
imaginary part of the ${\bf k}$-resolved $n=1$ image-state self-energy
$\Sigma(z,z';{\bf k}=0,\varepsilon_{\bf k})$, versus $z$, in the vicinity of
the (100) surfaces of Cu and Au, with use (in the case of the $G^0W$ and
$GW\Gamma$ approximations) of the adiabatic nonlocal xc kernel (ANLDA)
described in Sec.~\ref{fxc}. This figure shows that as occurs in the case of
the screened interaction xc effects partially compensate each other, leading
to an overall effect of no more than 5\% percent. We note that, as anticipated
in Ref.~\cite{maia2} for the case of Cu(111), although the ALDA leads to spurious results for the screened interaction our
more realistic ANLDA kernel yields self-energies that esentially coincide with
those obtained in the ALDA. 

\begin{table}
\caption{Reciprocal lifetimes, in linewidth units (meV), of the $n=1$ image
state (at $\bar\Gamma$, i.e., ${\bf k}=0$) on Cu(100), together with the most
recent measurement reported in
Ref.~\cite{hofer2}. All the single-particle wave functions and energies
entering Eqs.~(\ref{eq1}), (\ref{eq5}), (\ref{eq5p}), and (\ref{eq5pp}) have
been chosen to be the eigenfunctions and eigenvalues of the 1D Hamiltonian of
Chulkov {\it et al.}~\cite{chulkov1}. Effective masses for all the
single-particle energies entering Eqs.~(\ref{eq5}), (\ref{eq5p}), and
(\ref{eq5pp}) have been set equal to either the free-electron mass ($m_f=1$)
or to realistic values ($m_f\neq 1$). Various models have been considered for
the description of the electron self-energy, as obtained from (i)
Eq.~(\ref{eq5}) with the screened interaction of Eq.~(\ref{gh1}) (SRM), of
Eqs.~(\ref{wg})-(\ref{geta}) (PZ), and of Eqs.~(\ref{wsurf}) and
(\ref{eq:Xalda3}) with $f^{xc}[n_0](z,z';q,\omega)$ set equal to zero ($G^0W^0$), (ii) Eq.~(\ref{eq5p}) with the screened interaction of Eqs.~(\ref{wsurf}) and (\ref{eq:Xalda3}) and the ANLDA
xc kernel $f^{xc}[n_0](z,z';q,\omega)$ ($G^0W$), and (iii) Eq.~(\ref{eq5pp}) with
the effective screened interaction of Eqs.~(\ref{wsurf2}) and (\ref{eq:Xalda3})
and the ANLDA xc kernel $f^{xc}[n_0](z,z';q,\omega)$ ($GW\Gamma$).\label{tablecu1}}
\begin{ruledtabular}
\begin{tabular}{lllllll}
$m_f$ & Self-energy & Bulk & Vacuum & Mix & Total & Exp. \\
\hline
=1&SRM&18&3&-4&17&\\
=1&PZ&&55&&55&\\
=1&$G^0W^0$&24&14&-16&22&\\
$\neq 1$&$G^0W^0$&7&11.5&-1&{\bf 17.5}&\\
$\neq 1$&$G^0W$    &    &  & &24.5&\\
$\neq 1$&$GW\Gamma$&6.5&11.5&-1&{\bf 17}&\\
        &          &   &    &  &        &{\bf 16}\\
\end{tabular}
\end{ruledtabular}
\end{table}

\begin{table}
\caption{As in Table~\ref{tablecu1}, but for Cu(111) and together with the
reciprocal lifetime experimentally determined for this surface at low
temperature, $T=25\,{\rm K}$~\cite{knoesel}.\label{tablecu2}}
\begin{ruledtabular}
\begin{tabular}{lllllll}
$m_f$ & Self-energy & Bulk & Vacuum & Mix & Total & Exp. \\
\hline
=1&SRM&46&12&-22&36&\\
=1&PZ&&57&&57&\\
=1&$G^0W^0$&44&47&-54&37&\\
$\neq 1$&$G^0W^0$&32&34&-37&{\bf 29}&\\
$\neq 1$&$G^0W$     &  &   & &43&\\
$\neq 1$&$GW\Gamma$&30&35&-38&{\bf 28}&\\
        &          &   &    &  &        &{\bf 30}\\
\end{tabular}
\end{ruledtabular}
\end{table}
  
Tables~\ref{tablecu1} and \ref{tablecu2} exhibit the linewidth of the $n=1$
image state (at $\bar\Gamma$, i.e., ${\bf k}=0$) on the (100) and (111) surfaces of Cu, as obtained from Eq.~(\ref{eq1}) with (i) the image-state wave function of
Chulkov {\it et al.}~\cite{chulkov1} described in Sec.~\ref{sec:chulkov} and (ii) various approximations for the self-energy: $SRM$, $PZ$, $G^0W^0$, $G^0W$, and
$GW\Gamma$. Contributions to the
linewidth are separated as
follows
\begin{equation}
\Gamma_{ee}=\Gamma_{bulk}+\Gamma_{vac}+\Gamma_{mix},
\end{equation}
where $\Gamma_{bulk}$, $\Gamma_{vac}$, and $\Gamma_{mix}$ represent bulk,
vacuum, and mixed contributions, respectively, as obtained by confining the
integrals in Eqs.~(\ref{eq1}) to either bulk ($z<0,z'<0$), vacuum ($z>0,z'>0$),
or mixed ($z\ge 0,z'\le 0$ or $z\le 0,z'\ge 0$) coordinates.

First of all, we set all effective masses equal to the free-electron mass, and
focus on the role that an accurate description of the screened interaction
plays in the coupling of image states with the solid, by comparing to the
results obtained (within the $G^0W^0$ approximation) with the use of the SRM
screened interaction and (for the vacuum linewidth) the screened interaction of
Persson and Zaremba (PZ). We note that simplified jellium models for the
evaluation of the screened interaction yield unrealistic results for the
image-state lifetime. Bulk contributions to the linewidth are approximately
well described within the SRM, small differences
resulting from an approximate description, within this model, of the so-called
{\it begrenzung} or {\it boundary-effect} first described by Ritchie~\cite{ritchie1}. However, as quantum-mechanical details of the surface are ignored within this model, it fails to describe both vacuum and mixed contributions to the decay rate. These quantum-mechanical details
of the surface are approximately taken into account within the PZ approach, but
the PZ model cannot account for the
coupling of the image state with the crystal that occurs through the
penetration of the image-state wave function into the solid. Discrepancies
between vacuum contributions obtained in this model and in the more realistic
full $G^0W^0$ approach appear as a result of the PZ model being accurate
only for small ${\bf q}$ wave vectors and $\omega$ frequencies.

Now we account for the variation of the potential in the plane of the surface
through the introduction of a realistic effective mass for all surface and bulk
states. The effective masses of
the $n=1$ image state on Cu(100) and Cu(111) are close to the free-electron
mass ($m_i=1$). Nevertheless, the effective mass of the $n=0$ Shockley surface
state of Cu(111) and the unoccupied bulk states in Cu(111) and Cu(100), which
all contribute to the decay of the $n=1$ image state, considerably deviate from
the free-electron mass; Tables~\ref{tablecu1} and \ref{tablecu2} show that the
impact of this deviation on the $n=1$ image-state lifetime is not neglegible.

As for the impact of short-range xc effects, which are fully incorporated in
the framework of the $GW\Gamma$ approximation, Tables~\ref{tablecu1} and
\ref{tablecu2} show that the overall impact of these effects is small and
$GW\Gamma$ reciprocal lifetimes are close to their $G^0W^0$ counterparts.

\begin{table}
\caption{$G^{0}W^{0}$, $G^{0}W$, and $GW\Gamma$ reciprocal lifetimes, in
linewidth units (meV), of the $n=1$ image state (at $\bar\Gamma$, i.e., ${\bf k}=0$)
on Au(100). In the case of the $G^{0}W$ and $GW\Gamma$
reciprocal lifetimes, both ALDA and ANLDA xc kernels have been considered.}
\label{tableau}
\begin{ruledtabular}
\begin{tabular}{llll}
xc kernel & $G^{0}W^{0}$ & $G^{0}W$ & $GW\Gamma$\\
\hline
    &30&    &    \\
ALDA&    &42.5 &31\\
ANLDA&   &43 &31\\
\end{tabular}
\end{ruledtabular}
\end{table}

ALDA and ANLDA $GW\Gamma$ calculations of the reciprocal lifetimes of the $n=1$
image state on Au(100), never reported before, are exhibited in
Table~\ref{tableau}. For comparison, this Table also shows $G^0W^0$ and
$G^0W$ calculations, which in the case of the $G^0W$ have been obtained by
considering (as within the $GW\Gamma$ approximation) both local (ALDA) and
nonlocal (ANLDA) xc kernels. As occurs in the case of Cu~\cite{maia2}, the
results shown in Table~\ref{tableau} indicate that (i) a realistic adiabatic
nonlocal description of xc effects yields reciprocal lifetimes of image states
that esentially coincide with those obtained in the ALDA, and (ii) the overall
effect of short-range exchange and correlation is small, thereby $GW\Gamma$ reciprocal
lifetimes being close to their $G^0W^0$ counterparts.  

\begin{table}
\caption{Linewidth (inverse lifetime) of image states, in meV. The lifetime in
fs ($1\,{\rm fs}=10^{-15}s$) is obtained by noting that
$\hbar=658\,{\rm meV}\,{\rm fs}$. The numbers in brackets represent the
corresponding references. Electron-phonon linewidths are not included, since in the case of image states they are all expected to be below 1 meV~\cite{eiguren3}.}
\label{table4}
\begin{ruledtabular}
\begin{tabular}{llll}
&$G^0W^0$&$GW\Gamma$&Exp.\\
\hline
Li(110)& $37$~\cite{sarria} & &\\
Cu(100)& $17.5$~\cite{sarria} & $17$~\cite{maia2} &
$16.5\pm2.5$~\cite{hofer,shumay}\\
& & & 16~\cite{hofer2}\\
Cu(111)& $29$~\cite{sarria} & $28$~\cite{maia2} &
$30$~\cite{knoesel}\footnotemark[1]\\
       &                    &                   &
$29\pm 6$~\cite{weinelt}\footnotemark[2]\\
Ag(100)& 12~\cite{aran} & & $12\pm 1$~\cite{shumay} \\
Ag(111)& 36~\cite{aran} & & $21\pm 9/5$~\cite{lingle}\\
Au(100)& 30~\cite{chulkovchem}&31\footnotemark[3] & \\
Pd(111)& 30~\cite{pd}& &$26\pm 5/3$~\cite{pd}\\
Pt(111)& 23~\cite{pt}& &$25\pm 10/5$~\cite{pt}\\
Ni(100)& 33~\cite{chulkovchem}& &$41\pm 19/10$~\cite{ni100}\\
Ni(111)& 44~\cite{chulkovchem}& &$94\pm 71/28$~\cite{ni111}\\
Ru(1000)&47~\cite{echeap}&&59~\cite{ruexp}\\
\end{tabular}
\end{ruledtabular} 
\footnotetext[1]{At $T=25\,{\rm K}$.}
\footnotetext[2]{At $T=100\,{\rm K}$.}
\footnotetext[3]{This work.}
\end{table}

In Table~\ref{table4}, we compare self-consistent $G^0W^0$ and $GW\Gamma$ (as
obtained with the ANLDA xc kernel) calculations (with full inclusion of realistic
values of the effective mass of all bulk and surface states) with the existing
TR-2PPE data for the $n=1$ image state at the $\bar\Gamma$ point (${\bf k}=0$) on various simple, noble, and transition single-crystal surfaces. This table shows that
$G^0W^0$ and $GW\Gamma$ calculations are both in good agreement with TR-2PPE measurements except in the case of the (111) surfaces of the noble metal Ag and the transition metal Ni. The largest disagreement occurs in the case of Ni(111), where the $G^0W^0$ linewidth is smaller than the measured linewidth by approximately a factor of 2. This can be attributed to the necessity of a full {\it ab initio} description of the dynamical response of both $sp$ and $d$ electrons along the lines of Sec.~\ref{periodic}.

The role that occupied $d$ bands play in the dynamics of image states on silver surfaces was investigated in Ref.~\cite{aran} along the lines described in Sec.~\ref{delectrons}. It was concluded that $d$ electrons do play an important role as a consequence of the reduction (in the presence of $d$ electrons) of the surface-plasmon energy that allows the opening of a new decay channel. No surface-plasmon decay channel is opened, however, in the case of the other noble-metal surafces (Cu and Au), since in the presence of $d$ electrons the Cu and Au surface-plasmon energy is still too large.      

\subsection{Shockley states}

$G^0W^0$ calculations of the e-e inelastic lifetimes of excited holes at the surface-state band edge of the (111) surfaces of the noble metals Cu, Ag, and Au were first reported in Refs.~\cite{science} and \cite{echeap} within the 1D scheme of Chulkov {\it et al.}~\cite{chulkov1} (see Sec.~\ref{sec:chulkov}), accounting for the potential variation in the plane of the surface through the introduction of a realistic effective mass for the dispersion curve of both bulk and surface states. Within this model, however, all Shockley states have the same effective mass and the projected band structure is still inaccurate, especially at energies above the Fermi level,
as shown in Fig.~\ref{fig4} for Cu(111). As an alternative to the 1D model potential of Chulkov {\it et al.}~\cite{chulkov1}, Vergniory {\it et al.}~\cite{maia1} introduced the ${\bf k}$-dependent 1D potential of Eq.~(\ref{model}) that is constructed to reproduce the bulk energy bands and surface-state energy dispersion obtained from 3D first-principles calculations.

\begin{table}
\caption{$G^0W^0$, $G^0W$, and $GW\Gamma$ decay rates, in linewidth units (meV),
of an excited
hole at the band edge of the Shockley surface-state band of Cu(111)
($E_i=-0.44\,{\rm eV}$ and $k_i=0$). The $G^0W^0$ calculations have been performed either with the use of the ${\bf k}$-dependent 1D model potential of Eq.~(\ref{model}), as reported in Ref.~\cite{maia1}, or with the use of the 1D scheme of Chulkov
{\it et al.}~\cite{chulkov1}, as reported in Refs.~\cite{science} and \cite{echeap}. The $G^0W$ and $GW\Gamma$ calculations have been performed with the use of the 1D scheme of Chulkov {\it et al.}~\cite{chulkov1}, as reported in Ref.~\cite{maia2}. The experimental linewidth has been taken from the STM measurements reported in
Ref.~\cite{science}. The total decay rate $\Gamma_{\rm total}$ includes the e-ph decay rate of 7 meV reported in Ref.~\cite{eiguren2}. $\Gamma_{\rm inter}$ and
$\Gamma_{\rm intra}$ represent interband and intraband contributions to the e-e decay rate $\Gamma_{\rm e-e}$; these contributions come from the decay of the excited hole trhough the coupling with bulk states (interband contribution) or thorugh the coupling, within the surface-state band itself, with surface states of different wave vector
${\bf k}$ parallel to the surface (intraband contribution).}
\label{table6}
\begin{ruledtabular} \begin{tabular}{lllll}
&$\Gamma_{\rm inter}$&$\Gamma_{\rm intra}$ &$\Gamma_{\rm e-e}$&
$\Gamma_{\rm total}$\\
\hline
$G^0W^0$~\cite{maia1}&10&9&19&{\bf 26}\\
$G^0W^0$~\cite{science,echeap}&6&19&25&32\\
$G^0W$~\cite{maia2}& & &30.5&37.5\\
$GW\Gamma$~\cite{maia2}& & &24.5&31.5\\
Experiment& & & &{\bf 24}\\
\end{tabular}
\end{ruledtabular}
\end{table}

Table~\ref{table6} shows a comparison between the $G^0W^0$ calculations reported in Refs.~\cite{science,echeap} and \cite{maia1} for the inelastic lifetime of an excited hole at the band edge of the Shockley surface-state band of Cu(111), as obtained from Eqs.~(\ref{eq1}) and (\ref{eq2}) with the use of the 1D scheme of Chulkov {\it et al.}~\cite{chulkov1} and with the ${\bf k}$-dependent 1D model potential of
Eq.~(\ref{model}), respectively. The difference between the surface-state lifetime broadening of $25\,{\rm eV}$ reported in Refs.~\cite{science} and \cite{echeap} and the more accurate lifetime broadening of $19\,{\rm eV}$ reported in Ref.~\cite{maia1} is entirely due to a more accurate description in Ref.~\cite{maia1} of (i) the projected band structure and (ii) the wave-vector dependence of the surface-state wave functions entering the evaluation of the self-energy. $G^0W$ and $GW\Gamma$ calculations were reported in Ref.~\cite{maia2}, showing that as in the case of image states $GW\Gamma$ linewidths are only slightly lower than their $G^0W^0$ counterparts. 

At this point, we note that the linewidths of the Cu(111) Shockley state at
$\bar\Gamma$ based on the use of the two 1D models of Sec.~\ref{sec:chulkov} to describe the {\it initial} surface-state wave function (at $\bar\Gamma$) agree within less than 1 meV. The linewidths also agree within less than 1 meV when the actual surface-state dispersion (thick solid line of Fig.~\ref{fig4}) is replaced by the parabolic surface-state dispersion of the form dictated by the thin solid line of Fig.~\ref{fig4}. However, if one replaces in the calculation of Ref.~\cite{maia1} the wave-vector dependent surface-state orbitals obtained by solving a 1D Schr\"odinger equation with the potential of Eq.~(\ref{model}) by their less accurate counterparts used in Refs.~
\cite{science} and \cite{echeap}, the lifetime broadening is increased considerably (from 19 to 25 meV), showing the important role that the actual coupling between initial and final states plays in the surface-state decay mechanism.

\begin{table}
\caption{$G^{0}W^{0}$, $G^{0}W$, and $GW\Gamma$ reciprocal lifetimes, in
linewidth units (meV), of the $n=0$ Shockley state (at $\bar\Gamma$, i.e., ${\bf k}=0$)
on Au(111). In the case of the $G^{0}W$ and $GW\Gamma$
reciprocal lifetimes, both ALDA and ANLDA xc kernels have been considered.}
\label{tableaus}
\begin{ruledtabular}
\begin{tabular}{llll}
xc kernel & $G^{0}W^{0}$ & $G^{0}W$ & $GW\Gamma$\\
\hline
    &29&    &    \\
ALDA&    &39.5 &30\\
ANLDA&   &40 &30\\
\end{tabular}
\end{ruledtabular}
\end{table}

ALDA and ANLDA $GW\Gamma$ calculations of the reciprocal lifetimes of the Shockley surface state on Au(111), never reported before, are exhibited in
Table~\ref{tableaus}. For comparison, this Table also shows $G^0W^0$ and
$G^0W$ calculations, which in the case of the $G^0W$ have been obtained by
considering (as within the $GW\Gamma$ approximation) both local (ALDA) and
nonlocal (ANLDA) xc kernels. As occurs in the case of Cu~\cite{maia2}, the
results shown in Table~\ref{tableaus} indicate that (i) a realistic adiabatic
nonlocal description of xc effects yields reciprocal lifetimes of Shockley states
that esentially coincide with those obtained in the ALDA, and (ii) the overall
effect of short-range exchange and correlation is small, thereby $GW\Gamma$ reciprocal
lifetimes being close to their $G^0W^0$ counterparts.  

\begin{table}
\caption{Linewidth (inverse lifetime) of Shockley states, in meV. The lifetime in
fs ($1\,{\rm fs}=10^{-15}s$) is obtained by noting that
$\hbar=658\,{\rm meV}\,{\rm fs}$. The numbers in brackets represent the
corresponding references. The calculated values ($\Gamma_{calc}$) are decomposed into
e-e ($\Gamma_{e-e}$) and e-ph ($\Gamma_{e-ph}$) contributions. Since $GW\Gamma$ e-e linewidths are found to be very close to their $G^0W^0$ counterparts, only $G^0W^0$ calculations of the e-e linewidth are included here. In the case of Be(0001), calculations have been performed either with the use of the 1D scheme of Chulkov {\it et al.}~\cite{chulkov1}, as reported in Ref.~\cite{chulal}, or via a fully {\it ab initio} scheme along the lines of Sec.~\ref{periodic}, as reported in Ref.~\cite{chulbe}. In the case of Cu(111), calculations have been performed either with the use of the ${\bf k}$-dependent 1D model potential of Eq.~(\ref{model}), as reported in Ref.~\cite{maia1}, or with the use of the 1D scheme of Chulkov {\it et al.}~\cite{chulkov1}, as reported in
Refs.~\cite{science} and \cite{echeap}.\label{table8}}
\begin{ruledtabular}
\begin{tabular}{lllll}
&$\Gamma_{e-e}$&$\Gamma_{e-ph}$&$\Gamma_{calc}$&Exp.\\
\hline
Al(111)&  336~\cite{chulal}        & 36~\cite{chulal}  & 372 & $\sim 1500$~\cite{kevan1}
\footnotemark[1]\\
Mg(0001)& 83~\cite{chulal}         & 25~\cite{chulal}  & 108 & $\sim 500$~\cite{kevan1}
\footnotemark[1]\\
Be(0001)& 280~\cite{chulal}        & 80~\cite{chulbe}  & 360 &\\
        & 265~\cite{chulbe}        & 80~\cite{chulbe}  & 345 & 350~\cite{chulbe}\\
Cu(111)&  25~\cite{science,echeap} & 7~\cite{eiguren2} & 32  &                  \\
&         19~\cite{maia1}          & 7~\cite{eiguren2} & 26  & 24~\cite{science}\\
Ag(111)&   3~\cite{science,echeap} & 4~\cite{eiguren2} & 7   & 6.5~\cite{science}\\
Au(111)&  29~\cite{science,echeap} & 4~\cite{eiguren2} & 33  & 18~\cite{science}\\
\end{tabular}
\end{ruledtabular} 
\footnotetext[1]{At room temperature.}
\end{table}

The calculated and experimental linewidths of Shockley states at the $\bar\Gamma$ point of a variety of simple and noble metal surfaces are collected in Table~\ref{table8}. It had been argued in Ref.~\cite{campillo} that in the case of the noble metals deviations from electron dynamics in a free gas of $sp$ electrons due to the participation of $d$ electrons in the screening of e-e interactions are of key importance in the determination of the inelastic lifetime of bulk electronic states. Hence, Kliewer {\it et al.}~\cite{science} added this effect to the calculated $\Gamma_{e-e}$ following the approach originally suggested by Quinn~\cite{quinn}; they concluded that the screening of $d$ electrons reduces the e-e scattering considerably, thus improving the agreement with experiment. Nevertheless, it was shown in Ref.~\cite{aran} that in the case of Shockley states, whose decay is dominated by intraband transitions that are associated with very small values of the momentum transfer, the screening of $d$ electrons is expected to reduce the lifetime broadening only very slightly. Indeed, adding to the estimated Cu(100) Shockley $e-e$ linewidth at $\bar\Gamma$ reported recently in Ref.~\cite{maia1} (with no $d$-screening reduction) the e-ph linewidth of 7 meV reported in Ref.~\cite{eiguren2}, the calculated total linewidth is found to be $\Gamma_{calc}=26\,{\rm meV}$ in close agreement with the exprimentally measured linewidth of 24 meV, as shown in Table~\ref{table8}. An extension of the approach reported in Ref.~\cite{maia1} to the case of the other noble metals Ag and Au should yield calculated linewidths that are closer to experiment than those reported in Refs.~\cite{science} and \cite{echeap}.

The lifetime broadening of excited Shockley electrons beyond the $\bar\Gamma$ point (with ${\bf k}\neq 0$ and energies above the Fermi level - see Fig.~\ref{fig1}) was studied with the STM by B\"urgi {\it et al.}~\cite{stm2} on Cu(111) and by Vitali {\it et al.}~\cite{vitali} and Kliewer {\it et al.}~\cite{njp} on Ag(111). The
corresponding $G^0W^0$ calculations that follow the 1D scheme of Chulkov {\it et al.}~\cite{chulkov1} were reported in Refs.~\cite{ru} and \cite{vitali} for Cu(111) and Ag(111), respectively, but now accounting for the potential variation parallel to the surface by introducing not only a realistic effective mass for all bulk and surface states but also  surface-state orbitals that change with {\bf k} along the surface-state dispersion curve. More accurate calculations were later reported in the case of Cu(111)~\cite{maia1} with the use of the ${\bf k}$-dependent 1D model potential of Eq.~(\ref{model}), showing that the inelastic lifetimes of excited Shockley electrons happen to be very sensitive to the actual shape of the surface-state single-particle orbitals beyond the $\bar\Gamma$ point. A comparison between these more refined calculations and experiment demonstrated that there is close agreement at the surface-state band edge, i.e., at $\bar\Gamma$, as shown in Tables~\ref{table6} and \ref{table8}, and there is also reasonable agreement at low excitation energies above the Fermi level. At energies where the surface-state band merges into the continuum of bulk states, however, the calculated linewidths are found to be too low, which should be a signature of the need of a fully 3D description of the surface band structure.    

\section{Acknowledgments}

Partial support by the University of the Basque Country, the Basque
Unibertsitate eta Ikerketa Saila, the Spanish Ministerio de Educaci\'on
y Ciencia (Grant No. CSD2006-53), and the EC 6th framework Network of
Excellence NANOQUANTA (Grant No. NMP4-CT-2004-500198) are acknowledged. The authors
also wish to thank E. V. Chulkov, S. Crampin, P. M. Echenique, J. E. Inglesfield, and V. M. Silkin for enjoyable collaboration and discussions.


\begin{thebibliography}{1}
\bibitem{note1} This should be a reasonable approximation for the (100)
surfaces of the noble metlas, in which case the vacuum level is located near
the center of the band gap.
\bibitem{notearan} An exception is the case of Ag(100)~\cite{aran}. Due to the
presence of $d$-electron screening, this surface supports the excitation of
surface plasmons at a reduced energy of $\sim 3.7\,{\rm eV}$ that is just
below the $n=1$ image-state energy
$\varepsilon_{n=1}-\varepsilon_F=3.9\,{\rm eV}$.
\end{thebibliography}
\end{document}